\documentclass[11pt]{article}
\usepackage{amssymb}
\usepackage{amsmath}
\usepackage[all]{xy}
\input{epsf}
\usepackage{epsfig}
\usepackage{pifont}
\numberwithin{equation}{section}

\def\be{\begin{equation}}
\def\ee{\end{equation}}
\def\ba{\begin{align}}
\def\ea{\end{align}}
\def\beq{\begin{eqnarray}}
\def\eeq{\end{eqnarray}}

 \input{epsf}

 \usepackage{epsfig}

\begin{document}

\title{\Large{\bf 
The Hirota equation for string theory in $AdS_5\times S^5$ from the fusion of line operators}} 
\author{Raphael Benichou}
%\date{}
\maketitle
\begin{center}
 Theoretische Natuurkunde, Vrije Universiteit Brussel and \\
The International Solvay Institutes,\\
 Pleinlaan 2, B-1050 Brussels, Belgium \\
 \textsl{raphael.benichou@vub.ac.be}
\end{center}

 \begin{abstract}
We present a perturbative derivation of the T-system that is believed to encode the exact spectrum of planar $\mathcal{N}=4$ SYM.
The T-system is understood as an operator identity between some special line operators, the quantum transfer matrices.
By computing the quantum corrections in the process of fusion of transfer matrices, we show that the T-system holds up to first order in a semi-classical expansion. This derivation does not rely on any assumption.
We also discuss the extension of the proof to other theories, including models describing string theory on various AdS spaces.
\end{abstract}

%\newpage

\section{Introduction}

The AdS/CFT correspondence \cite{Maldacena:1997re}\cite{Gubser:1998bc}\cite{Witten:1998qj} states that type IIB string theory in $AdS_5 \times S^5$ is dual to $\mathcal{N}=4$ SYM with gauge group $SU(N)$. In the planar limit of the gauge theory, or equivalently in the classical limit of the string theory, integrable structures appear. This observation lead to impressive progress in the understanding of this system (see \cite{Beisert:2010jr} for a review). In particular a solution, known as the Y-system, has been put forward to solve the spectrum problem \cite{Gromov:2009tv}. This solution takes the form of an infinite set of equations for the so-called Y-functions. For each state in the theory, there is one solution to this set of equations. The energy of the state can be easily computed once the corresponding solution is known.
There is by now solid evidence in favor of the validity of the Y-system. The most impressive success was the correct prediction of subleading terms in the dimension of the Konishi operator both at large \cite{Gromov:2009zb} and at small \cite{Arutyunov:2010gb} 't Hooft coupling.

The Y-system can be derived using the Thermodynamic Bethe Ansatz approach \cite{Zamolodchikov:1989cf}\cite{Bajnok:2010ke}. This was achieved in \cite{Gromov:2009bc}\cite{Bombardelli:2009ns}\cite{Arutyunov:2009ur}. This approach was very successful. However it relies on several assumptions that are notoriously difficult to prove. Firstly one has to assume quantum integrability to begin with. Then one has to formulate the string hypothesis. Finally this approach only provides the ground state energy; the reason why the energy of the excited states can be obtained by analytic continuation is not understood. In this work, we present a new approach to derive the Y-system from first principles. We will use elementary tools of two-dimensional conformal field theory. The approach presented here is somehow similar in spirit to the seminal work of \cite{Bazhanov:1994ft} where the Y-system was derived for minimal models.

\section{Strategy of the proof}

\begin{figure}[b]
\includegraphics[width=\linewidth]{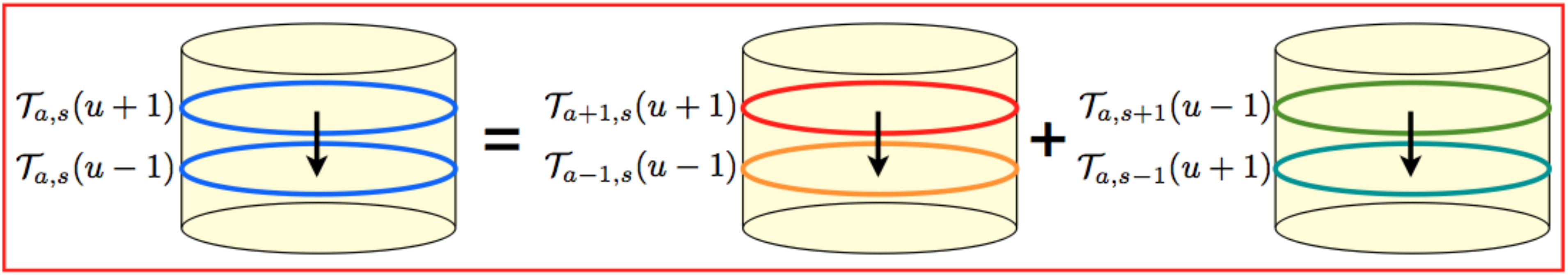}
\caption{The T-system understood as an operator identity between transfer matrices associated to different representations. The product in the T-system is understood as the fusion of line operators.}
\label{fig:1}
\end{figure}

Let us explain the strategy we will use to derive the Y-system.
Up to a change of variables, the Y-system can be rewritten as a T-system, or Hirota equation:
\begin{equation}\label{Tsystem} \mathcal{T}_{a,s}(u + 1) \mathcal{T}_{a,s}(u - 1) = 
  \mathcal{T}_{a+1,s}(u+1)\mathcal{T}_{a-1,s}(u-1) +  \mathcal{T}_{a,s+1}(u-1)\mathcal{T}_{a,s-1}(u+1) \end{equation}
The T-functions have two integer labels $a$ and $s$ that take values in a T-shaped lattice \cite{Gromov:2010kf}\cite{Volin:2010xz}. The integers $(a,s)$ can be understood as giving the size of a rectangular Young tableau. Thus each T-function is naturally associated to a representation of the global symmetry group $PSU(2,2|4)$. The T-functions also depend on a spectral parameter $u$.

It is believed that the T-functions are related to the transfer matrices of the worldsheet theory (see e.g. \cite{Gromov:2009tq}\cite{Gromov:2010vb}).
The transfer matrices are defined as:
\begin{equation}\label{defT} \mathcal{T}_R(u) = STr \ P \exp\left(-\oint A_R(u)\right) \end{equation}
where $R$ labels a representation of $PSU(2,2|4)$ and $A_R(u)$ is a flat connection that transforms in the representation $R$.
Notice that the classical transfer matrix is the supertrace of a group element. Thus it is a supercharacter. It is known that supercharacters of $PSU(2,2|4)$ satisfy the following identity \cite{Kazakov:2007na}:
\begin{equation}\label{charId1} \chi_{(a,s)}^2 = \chi_{(a+1,s)}\chi_{(a-1,s)}+\chi_{(a,s+1)}\chi_{(a,s-1)} \end{equation}
This character identity is nothing but the T-system \eqref{Tsystem} when the shifts of the spectral parameter are neglected.
This suggests that the shifts in the T-system come from some kind of quantum effects. The remaining question is to identify the relevant quantum effects.
In this work we propose an answer to this question. First we take the identification between the T-functions and the transfer matrices seriously, thus promoting the T-system \eqref{Tsystem} to an operator identity. We postulate that the operator product appearing in the T-system is the fusion of line operator, that is the process of bringing the integration contours on top of each other (see Figure \ref{fig:1}). Then the shifts of the spectral parameter in the T-system \eqref{Tsystem} should come from quantum effects in the process of fusion. We will demonstrate that this is the case up to first order in the large 't Hooft coupling expansion. More precisely, we will compute the leading quantum correction in the process of fusion of the transfer matrices, and show that it leads to the correct shifts in the T-system.
The details of the computations described below can be found in \cite{Benichou:2011ch}.

\section{Line operators in the pure spinor string on $AdS_5 \times S^5$}

To describe string theory in $AdS_5 \times S^5$, we use the pure spinor formalism \cite{Berkovits:2000fe}\cite{Mazzucato:2011jt}.
This theory admits a one parameter family of flat connections \cite{Bena:2003wd}\cite{Vallilo:2003nx} denoted by $A_R(u)$, which can be written as a linear combination of the elementary currents of the model. Using these flat connections we can define the transfer matrix \eqref{defT}.

One has to be careful when working with line operators in a quantum theory. Indeed the collisions between integrated operators generically lead to UV divergences that have to be regularized, and then cancelled by proper renormalization of the line operators \cite{Bachas:2004sy}. We use a principal value regularization scheme to deal with such divergences \cite{Mikhailov:2007eg}. Remarkably, the divergences in the transfer matrix do cancel in our case, at least at leading order in perturbation theory \cite{Mikhailov:2007mr}\cite{Benichou:2011ch}. This properties essentially follows from the vanishing of the dual Coxeter number of the supergroup $PSU(2,2|4)$. An important consequence is that the quantum transfer matrix does not need to be renormalized.

Let us now consider the computation of the quantum corrections in the process of fusion of line operators. The quantum corrections come from the OPEs between the integrated operators on different contours. Actually only the part of the OPE that is anti-symmetric under the exchange of the two operators contributes  \cite{Mikhailov:2007eg}\cite{Benichou:2011ch}. The symmetric part of the OPE is understood as being a quantum dressing of the line operator  obtained once the fusion has been performed. The anti-symmetric OPE between two connections, that is the commutator of equal-time connections, takes the canonical form of a $(r,s)$ system \cite{Magro:2008dv}\cite{Vicedo:2009sn}\cite{Vicedo:2010qd}\cite{Benichou:2011ch} of the type introduced by Maillet \cite{Maillet:1985fn}\cite{Maillet:1985ek}:
\begin{align}\label{r,sSyst}
[ A_R(u;\sigma), A_{R'}(u';\sigma')] = &2s \partial_\sigma\delta^{(2)}(\sigma-\sigma') + [ A_R(u;\sigma)+ A_{R'}(u';\sigma'),r] \delta^{(2)}(\sigma-\sigma') \cr
& + [ A_R(u;\sigma)- A_{R'}(u';\sigma'),s] \delta^{(2)}(\sigma-\sigma')
\end{align}
where $r$ and $s$ are come constant matrices transforming in the tensor product of representations $R \otimes R'$.

It turns out that the fusion of transfer matrices is trivial at first order in perturbation theory. This implies in particular that the transfer matrices do commute at this order. To obtain the leading quantum correction in the fusion of transfer matrices, we need to go up to second order in perturbation theory. The  result for the quantum corrections is rather complicated and the explicit expressions as well as the details of the computations can be found in \cite{Benichou:2011ch}.
Luckily this complicated result simplifies in a particular semi-classical limit. Let $\mathcal{T}_R(u)$ and $\mathcal{T}_{R'}(u')$ be the two transfer matrices that we are fusing. We consider the limit $u\gg1$, $u' \gg 1$ and $u-u'\ll u,u'$. 
This limit is the one that is relevant for our initial purpose of deriving the T-system \eqref{Tsystem}. Indeed remember that the semi-classical limit of this system was identified as the limit where the shifts of the spectral parameter are small.
In this limit the leading quantum correction in the process of fusion is essentially equal to the line operator $\mathcal{T}_R(u) \mathcal{T}_{R'}(u')$ but with an additional operator $\tilde K$ integrated in between the various connections. This additional integrated operator $\tilde K$ takes the schematic form:
\begin{equation}\label{tK} \tilde K \sim \# \partial_u A^a(u) {f_a}^{bc} {f_c}^{de} t_b t_d t_e \end{equation}
where $\#$ is a coefficient that depends only on the spectral parameters and on the coupling constant, $a,b,c,d,e$ are adjoint indices, the $f$'s are structure constants and the $t$'s are generators of the Lie superalgebra $psu(2,2|4)$.

\section{Derivation of the Hirota equation}

Using the computation of the leading quantum corrections in the process of fusion, we can now prove the validity of the T-system \eqref{Tsystem} up to first order in the semi-classical expansion.
We start from the T-system \eqref{Tsystem}, in which the T-functions are now understood as quantum transfer matrices and the product is understood as the fusion. We perform a semi-classical expansion, assuming that the shifts of the spectral parameter are small:
\begin{align}\label{Tsyst1} \sum_{R,R'} & \mathcal{T}_{R}(u+1) \mathcal{T}_{R'}(u-1) =  \sum_{R,R'} \mathcal{T}_{R}(u) \mathcal{T}_{R'}(u) \cr
&+  \sum_{R,R'} \left(  \partial_u \mathcal{T}_{R}(u) \mathcal{T}_{R'}(u) -  \mathcal{T}_{R}(u) \partial_u \mathcal{T}_{R'}(u)\right)
+ \left( \begin{array}{c} \mathrm{Quantum\ corrections}\\ \mathrm{ from\ fusion} \end{array}\right)  + ...
\end{align}
In order to simplify the writing we denoted by $\sum_{R,R'}$ the sum over representations that appears in the T-system \eqref{Tsystem}.
The T-system is valid if the left-hand side (or equivalently the right-hand side) of \eqref{Tsyst1} vanishes. 
The first term on the right-hand side of \eqref{Tsyst1} is the classical term. It vanishes thanks to the character identity \eqref{charId1}.
The second term on the right-hand side of \eqref{Tsyst1} comes from the derivative expansion. It does not vanish on its own, so it has to cancel against the leading quantum correction coming from fusion.
To show that this is the case, we have to use some character identities that were derived in \cite{Kazakov:2007na}. These identities are valid for the particular combination of representations that appears in the T-system. Thanks to these identities, we can essentially replace the complicated contraction of structure constants and generators that appears in the operators $\tilde K$ \eqref{tK} by a single generator: $ {f_a}^{bc} {f_c}^{de} t_b t_d t_e \to t_a$. This implies that the operator $\tilde K$ itself can be replaced by the derivative of the flat connection with respect to the spectral parameter. Consequently the leading quantum correction coming from fusion can be written in terms of derivative of transfer matrices. After a careful treatment of all numerical factors \cite{Benichou:2011ch}, we find that the leading quantum correction from fusion exactly cancels against the second term in the right-hand side of \eqref{Tsyst1}. This completes the derivation of the T-system from first-principles up to first order in the semi-classical expansion.

\section{Generalizations and conclusion}

Let us first summarize the main results presented here. The main technical result is the computation of the quantum corrections in the fusion of line operators in the pure spinor string on $AdS_5 \times S^5$, up to second order in perturbation theory. Then we used this result to prove the validity of the T-system up to first order in the semi-classical expansion.

Let us briefly compare the approach presented here with the Thermodynamic Bethe Ansatz approach. The derivation of the T-system that we obtained  has two major advantages. Firstly, it does not rely on any hypothesis. Secondly, since we derived the T-system as an operator identity, it is clear that it holds for all states in the theory and not only for the ground state. On the other hand, the TBA approach also has some advantages. In particular it gives the full T-system in one go, and not only up to some order in perturbation theory. It also provides a formula to extract the energy from the T-functions, and gives some indications about the analytic properties of these functions.
Presumably the last two points can be addressed with the elementary tools of conformal field theory that we used in this work. It would be very interesting to investigate these points.

An important question is whether the derivation of the Hirota equation that we presented here can be generalized to other models. There is at least one family of models where this is the case: the non-linear sigma-models on the supergroup $PSl(n|n)$. This was shown in \cite{Benichou:2010ts}, building up on earlier studies of these models \cite{Ashok:2009xx}\cite{Benichou:2010rk}.
This result is relevant for string theory in $AdS_3 \times S^3$, supported by RR and/or NS fluxes. Indeed in this background perturbative string theory can be defined in the hybrid formalism as the sigma model on $PSU(1,1|2)$ coupled to ghosts \cite{Berkovits:1999im}. Thus the results of \cite{Benichou:2010ts} imply that the Y-system is also realized in string theory in $AdS_3 \times S^3$, at least up to first order in the semi-classical expansion.

A close look at the computations allows us to characterize rather precisely the set of models for which the derivation of the T-system applies. There are essentially two features that are almost sufficient for the proof to work. The first one is that the model admits a one-parameter family of flat connections,  which commutator can be written as a canonical $(r,s)$ system. The second one is that the global symmetry group of the model has vanishing dual Coxeter number. Consequently the chances are good that the proof can be easily generalized for string theory on $AdS_4 \times CP^3$, $AdS_2 \times S^2$, and essentially all string backgrounds that were identified in \cite{Zarembo:2010sg}.

\section*{Acknowledgments}
  The author is a Postdoctoral researcher of FWO-Vlaanderen.
This research is supported in part by the Belgian Federal Science Policy Office through the Interuniversity Attraction Pole IAP VI/11 and by FWO-Vlaanderen through project G011410N.


\begin{thebibliography}{[1]}

  %\cite{Maldacena:1997re}
\bibitem{Maldacena:1997re}
  J.~M.~Maldacena,
 % ``The large N limit of superconformal field theories and supergravity,''
  Adv.\ Theor.\ Math.\ Phys.\  {\bf 2} (1998) 231
  [Int.\ J.\ Theor.\ Phys.\  {\bf 38} (1999) 1113]
  %[arXiv:hep-th/9711200].
  %%CITATION = IJTPB,38,1113;%%
  
%\cite{Gubser:1998bc}
\bibitem{Gubser:1998bc}
  S.~S.~Gubser, I.~R.~Klebanov, A.~M.~Polyakov,
  %``Gauge theory correlators from noncritical string theory,''
  Phys.\ Lett.\  {\bf B428 } (1998)  105-114.
  %[hep-th/9802109].
  
%\cite{Witten:1998qj}
\bibitem{Witten:1998qj}
  E.~Witten,
 % ``Anti-de Sitter space and holography,''
  Adv.\ Theor.\ Math.\ Phys.\  {\bf 2 } (1998)  253-291.
 % [hep-th/9802150]

%\cite{Beisert:2010jr}
\bibitem{Beisert:2010jr}
  N.~Beisert, C.~Ahn, L.~F.~Alday, Z.~Bajnok, J.~M.~Drummond, L.~Freyhult, N.~Gromov, R.~A.~Janik {\it et al.},
  %``Review of AdS/CFT Integrability: An Overview,''
  [arXiv:1012.3982 [hep-th]]  

%\cite{Gromov:2009tv}
\bibitem{Gromov:2009tv}
  N.~Gromov, V.~Kazakov and P.~Vieira,
 % ``Exact Spectrum of Anomalous Dimensions of Planar N=4 Supersymmetric
 % Yang-Mills Theory,''
  Phys.\ Rev.\ Lett.\  {\bf 103}, 131601 (2009)
  %[arXiv:0901.3753 [hep-th]].
  %%CITATION = PRLTA,103,131601;%%

%\cite{Gromov:2009zb}
\bibitem{Gromov:2009zb}
  N.~Gromov, V.~Kazakov, P.~Vieira,
 % ``Exact Spectrum of Planar ${\cal N}=4$ Supersymmetric Yang-Mills Theory: Konishi Dimension at Any Coupling,''
  Phys.\ Rev.\ Lett.\  {\bf 104 } (2010)  211601.
  %[arXiv:0906.4240 [hep-th]].
  
%\cite{Arutyunov:2010gb}
\bibitem{Arutyunov:2010gb}
  G.~Arutyunov, S.~Frolov, R.~Suzuki,
 % ``Five-loop Konishi from the Mirror TBA,''
  JHEP {\bf 1004 } (2010)  069.
%  [arXiv:1002.1711 [hep-th]].

%\cite{Zamolodchikov:1989cf}
\bibitem{Zamolodchikov:1989cf}
  A.~B.~Zamolodchikov,
 % ``Thermodynamic Bethe Ansatz In Relativistic Models. Scaling Three State Potts And Lee-yang Models,''
  Nucl.\ Phys.\  {\bf B342}, 695-720 (1990).
  
%\cite{Bajnok:2010ke}
\bibitem{Bajnok:2010ke}
  Z.~Bajnok,
%  ``Review of AdS/CFT Integrability, Chapter III.6: Thermodynamic Bethe Ansatz,''
  [arXiv:1012.3995 [hep-th]]  
  
%\cite{Gromov:2009bc}
\bibitem{Gromov:2009bc}
  N.~Gromov, V.~Kazakov, A.~Kozak, P.~Vieira,
 % ``Exact Spectrum of Anomalous Dimensions of Planar N = 4 Supersymmetric Yang-Mills Theory: TBA and excited states,''
  Lett.\ Math.\ Phys.\  {\bf 91 } (2010)  265-287.
%  [arXiv:0902.4458 [hep-th]].
  
%\cite{Bombardelli:2009ns}
\bibitem{Bombardelli:2009ns}
  D.~Bombardelli, D.~Fioravanti, R.~Tateo,
%  ``Thermodynamic Bethe Ansatz for planar AdS/CFT: A Proposal,''
  J.\ Phys.\ A {\bf A42 } (2009)  375401.
%  [arXiv:0902.3930 [hep-th]].
  
%\cite{Arutyunov:2009ur}
\bibitem{Arutyunov:2009ur}
  G.~Arutyunov, S.~Frolov,
%  ``Thermodynamic Bethe Ansatz for the AdS(5) x S(5) Mirror Model,''
  JHEP {\bf 0905 } (2009)  068.
%  [arXiv:0903.0141 [hep-th]]. 

%\cite{Bazhanov:1994ft}
\bibitem{Bazhanov:1994ft}
  V.~V.~Bazhanov, S.~L.~Lukyanov, A.~B.~Zamolodchikov,
%  ``Integrable structure of conformal field theory, quantum KdV theory and thermodynamic Bethe ansatz,''
  Commun.\ Math.\ Phys.\  {\bf 177 } (1996)  381-398.
  %[hep-th/9412229]
  
%\cite{Gromov:2010kf}
\bibitem{Gromov:2010kf}
  N.~Gromov, V.~Kazakov,
%  ``Review of AdS/CFT Integrability, Chapter III.7: Hirota Dynamics for Quantum Integrability,''
  [arXiv:1012.3996 [hep-th]].
  
%\cite{Volin:2010xz}
\bibitem{Volin:2010xz}
  D.~Volin,
  %``String hypothesis for gl(n|m) spin chains: a particle/hole democracy,''
  [arXiv:1012.3454 [hep-th]].  
  
%\cite{Gromov:2009tq}
\bibitem{Gromov:2009tq}
  N.~Gromov,
%  ``Y-system and Quasi-Classical Strings,''
  JHEP {\bf 1001 } (2010)  112.
  %[arXiv:0910.3608 [hep-th]].
  
%\cite{Gromov:2010vb}
\bibitem{Gromov:2010vb}
  N.~Gromov, V.~Kazakov, Z.~Tsuboi,
  %``PSU(2,2|4) Character of Quasiclassical AdS/CFT,''
  JHEP {\bf 1007 } (2010)  097.
%  [arXiv:1002.3981 [hep-th]].

%\cite{Kazakov:2007na}
\bibitem{Kazakov:2007na}
  V.~Kazakov and P.~Vieira,
 % ``From Characters to Quantum (Super)Spin Chains via Fusion,''
  JHEP {\bf 0810} (2008) 050
  %[arXiv:0711.2470 [hep-th]].
  %%CITATION = JHEPA,0810,050;%%
  
%\cite{Benichou:2011ch}
\bibitem{Benichou:2011ch}
  R.~Benichou,
  %``First-principles derivation of the AdS/CFT Y-systems,''
  JHEP {\bf 1110} (2011) 112
  %[arXiv:1108.4927 [hep-th]].
  %%CITATION = ARXIV:1108.4927;%%
  
%\cite{Berkovits:2000fe}
\bibitem{Berkovits:2000fe}
  N.~Berkovits,
 % ``Super-Poincare covariant quantization of the superstring,''
  JHEP {\bf 0004} (2000) 018
%  [arXiv:hep-th/0001035].
  %%CITATION = JHEPA,0004,018;%%
  
%\cite{Mazzucato:2011jt}
\bibitem{Mazzucato:2011jt}
  L.~Mazzucato,
 % ``Superstrings in AdS,''
  [arXiv:1104.2604 [hep-th]].

%\cite{Bena:2003wd}
\bibitem{Bena:2003wd}
  I.~Bena, J.~Polchinski and R.~Roiban,
  %``Hidden symmetries of the AdS(5) x S**5 superstring,''
  Phys.\ Rev.\  D {\bf 69} (2004) 046002
  %[arXiv:hep-th/0305116].
  %%CITATION = PHRVA,D69,046002;%%

%\cite{Vallilo:2003nx}
\bibitem{Vallilo:2003nx}
  B.~C.~Vallilo,
  %``Flat currents in the classical AdS(5) x S**5 pure spinor superstring,''
  JHEP {\bf 0403} (2004) 037
  %[arXiv:hep-th/0307018].
  %%CITATION = JHEPA,0403,037;%%
  
  %\cite{Bachas:2004sy}
\bibitem{Bachas:2004sy}
  C.~Bachas, M.~Gaberdiel,
  %``Loop operators and the Kondo problem,''
  JHEP {\bf 0411 } (2004)  065.
  %[hep-th/0411067].
  
    %\cite{Mikhailov:2007eg}
\bibitem{Mikhailov:2007eg}
  A.~Mikhailov, S.~Schafer-Nameki,
 % ``Algebra of transfer-matrices and Yang-Baxter equations on the string worldsheet in AdS(5) x S(5),''
  Nucl.\ Phys.\  {\bf B802 } (2008)  1-39.
  %[arXiv:0712.4278 [hep-th]].
  
%\cite{Mikhailov:2007mr}
\bibitem{Mikhailov:2007mr}
  A.~Mikhailov, S.~Schafer-Nameki,
%  ``Perturbative study of the transfer matrix on the string worldsheet in AdS(5) x S**5,''
  [arXiv:0706.1525 [hep-th]].    

    %\cite{Magro:2008dv}
\bibitem{Magro:2008dv}
  M.~Magro,
%  ``The Classical Exchange Algebra of AdS(5) x S**5,''
  JHEP {\bf 0901 } (2009)  021.
  %[arXiv:0810.4136 [hep-th]].
  
%\cite{Vicedo:2009sn}
\bibitem{Vicedo:2009sn}
  B.~Vicedo,
 % ``Hamiltonian dynamics and the hidden symmetries of the AdS(5) x S**5 superstring,''
  JHEP {\bf 1001}, 102 (2010).
  %[arXiv:0910.0221 [hep-th]].

%\cite{Vicedo:2010qd}
\bibitem{Vicedo:2010qd}
  B.~Vicedo,
%  ``The classical R-matrix of AdS/CFT and its Lie dialgebra structure,''
  Lett.\ Math.\ Phys.\  {\bf 95 } (2011)  249-274.
  %[arXiv:1003.1192 [hep-th]].

  %\cite{Maillet:1985fn}
\bibitem{Maillet:1985fn}
  J.~M.~Maillet,
%  ``Kac-moody Algebra And Extended Yang-baxter Relations In The O(n) Nonlinear Sigma Model,''
  Phys.\ Lett.\  {\bf B162 } (1985)  137.
  
  %\cite{Maillet:1985ek}
\bibitem{Maillet:1985ek}
  J.~M.~Maillet,
  %``New Integrable Canonical Structures In Two-dimensional Models,''
  Nucl.\ Phys.\  {\bf B269 } (1986)  54.  
  
  %\cite{Benichou:2010ts}
\bibitem{Benichou:2010ts}
  R.~Benichou,
 % ``Fusion of line operators in conformal sigma-models on supergroups, and the Hirota equation,''
  JHEP {\bf 1101 } (2011)  066.
  %[arXiv:1011.3158 [hep-th]] 

%\cite{Ashok:2009xx}
\bibitem{Ashok:2009xx}
  S.~K.~Ashok, R.~Benichou and J.~Troost,
  %``Conformal Current Algebra in Two Dimensions,''
  JHEP {\bf 0906} (2009) 017
 % [arXiv:0903.4277 [hep-th]].
  %%CITATION = ARXIV:0903.4277;%%

%\cite{Benichou:2010rk}
\bibitem{Benichou:2010rk}
  R.~Benichou and J.~Troost,
  %``The Conformal Current Algebra on Supergroups with Applications to the Spectrum and Integrability,''
  JHEP {\bf 1004} (2010) 121
 % [arXiv:1002.3712 [hep-th]].
  %%CITATION = ARXIV:1002.3712;%%
  
%\cite{Berkovits:1999im}
\bibitem{Berkovits:1999im}
  N.~Berkovits, C.~Vafa and E.~Witten,
 % ``Conformal field theory of AdS background with Ramond-Ramond flux,''
  JHEP {\bf 9903} (1999) 018
  %[arXiv:hep-th/9902098].
  %%CITATION = JHEPA,9903,018;%%  

%\cite{Zarembo:2010sg}
\bibitem{Zarembo:2010sg}
  K.~Zarembo,
  %``Strings on Semisymmetric Superspaces,''
  JHEP {\bf 1005} (2010) 002
  %[arXiv:1003.0465 [hep-th]].
  %%CITATION = ARXIV:1003.0465;%%
 
  

\end{thebibliography}
\end{document}